\documentclass[runningheads]{llncs}
\usepackage{xcolor}

\usepackage{graphicx}
\usepackage{url}
\usepackage{soul}
\usepackage{graphicx} 
\usepackage{siunitx}
\usepackage{verbatim}
\usepackage[misc]{ifsym}
\usepackage{amsmath}
\usepackage{pifont}
\usepackage{algorithm2e}
\usepackage{caption}
\usepackage{subcaption}
\usepackage[labelformat=simple]{subcaption}

\usepackage{multirow,multicol,ctable}
\usepackage{amsmath,amssymb,amsfonts}
\usepackage{algorithmic}
\usepackage{textcomp} 
\usepackage{booktabs}
\usepackage{subcaption}
\usepackage{amsfonts}
\usepackage{multirow,multicol,ctable,arydshln}
\usepackage{float}
\usepackage{makecell}
\usepackage{color}
\usepackage[marginal]{footmisc}

\begin{document}
\title{Cross-Phase Mutual Learning Framework for Pulmonary Embolism Identification on Non-Contrast CT Scans}
\titlerunning{CPMN for Pulmonary Embolism Identification on Non-Contrast CT Scans}
%
\author{Bizhe Bai\inst{1,2,3,*} \and Yan-Jie Zhou\inst{1,2,4,*} \and Yujian Hu \inst{5} \and Tony C. W. Mok \inst{1,2} \and \\ Yilang Xiang \inst{5} \and Le Lu \inst{1} \and Hongkun Zhang\inst{5} \and Minfeng Xu\inst{1,2}\textsuperscript{(\Letter)} }
\authorrunning{B.Z. Bai \textit{et al.}}

\institute{DAMO Academy, Alibaba Group \and
Hupan Lab, Hangzhou, China \\
\email{eric.xmf@alibaba-inc.com} \and
School of Information Science and Technology, Fudan University, Shanghai, China \and
College of Computer Science and Technology, Zhejiang University, Hangzhou, China \and
Department of Vascular Surgery, The First Affiliated Hospital of Zhejiang University School of Medicine, Hangzhou, China
}
%
%
\maketitle              
\footnote{*These authors contributed equally to this work.\\}
\begin{abstract}
Pulmonary embolism (PE) is a life-threatening condition where rapid and accurate diagnosis is imperative yet difficult due to predominantly atypical symptomatology. Computed tomography pulmonary angiography (CTPA) is acknowledged as the gold standard imaging tool in clinics, yet it can be contraindicated for emergency department (ED) patients and represents an onerous procedure, thus necessitating PE identification through non-contrast CT (NCT) scans. In this work, we explore the feasibility of applying a deep-learning approach to NCT scans for PE identification. We propose a novel Cross-Phase Mutual learNing framework (CPMN) that fosters knowledge transfer from CTPA to NCT, while concurrently conducting embolism segmentation and abnormality classification in a multi-task manner. The proposed CPMN leverages the Inter-Feature Alignment (IFA) strategy that enhances spatial contiguity and mutual learning between the dual-pathway network, while the Intra-Feature Discrepancy (IFD) strategy can facilitate precise segmentation of PE against complex backgrounds for single-pathway networks. For a comprehensive assessment of the proposed approach, a large-scale dual-phase dataset containing 334 PE patients and 1,105 normal subjects has been established. Experimental results demonstrate that CPMN achieves the leading identification performance, which is 95.4\% and 99.6\% in patient-level sensitivity and specificity on NCT scans, indicating the potential of our approach as an economical, accessible, and precise tool for PE identification in clinical practice.

\end{abstract}

\section{Introduction}

Pulmonary embolism (PE) is a critical and potentially lethal pulmonary condition, which occupies the third position in severity, trailing only behind myocardial infarction and sudden cardiac death \cite{turetz2018epidemiology}. It typically arises from a thrombotic event within the deep venous network of the lower limbs, which subsequently embarks on a path through the bloodstream, advances to the cardiac region, and culminates in an obstruction within the pulmonary arterial network \cite{lapner2013diagnosis}. The predominant factor contributing to preventable mortality in PE cases is not a therapeutic shortfall, but rather the omission of accurate diagnosis \cite{fedullo2003evaluation}. 

Within the realm of PE diagnostic modalities, computed tomographic pulmonary angiography (CTPA) has emerged as the gold standard imaging tool, facilitating visualization of pulmonary filling defects through the utilization of contrast agents \cite{abdellatif2021diagnostic}. Unfortunately, certain patient populations within emergency departments (ED) are unable to easily undergo intravenous contrast-enhanced computed tomography scans, predominantly attributable to renal impairment or hypersensitivity to iodine-based contrast agents \cite{thom2017never}. On the contrary, non-contrast computed tomography (NCT) can be performed within seconds and is an economical and accessible tool. Nevertheless, the assessment of NCT scans by radiologists lacks the requisite sensitivity and specificity to dependably diagnose PE \cite{sun2014detection}. Therefore, developing an automatic and accurate PE identification framework on NCT scans is of paramount importance.

With recent advances in deep learning, an increasing number of researchers have devoted efforts to developing automated algorithms for PE identification on CTPA scans \cite{huang2020penet,yuan2021resd,chen2024scunet++}. Huang \textit{et al.} \cite{huang2020penet} presented a 3D convolutional neural network (CNN) for PE identification by decoupling the issue as a classification task, yet it lacks the capability to furnish precise localization. Recently, Yuan \textit{et al.} \cite{yuan2021resd} proposed a ResD-UNet framework for pulmonary artery segmentation, which enhances accuracy and efficiency through the integration of the U-Net architecture with innovative residual-dense blocks and a composite loss function, thereby tackling the challenge in assessing the severity of PE. Chen \textit{et al.} \cite{chen2024scunet++} introduced an automated segmentation approach for PE, termed SCUNet++, which integrates the strengths of UNet++, multiple fusion dense skip connections, the Swin-Transformer attention mechanism, and the Swin-UNet architecture. Conversely, the realm of PE identification on NCT scans remains comparatively underexplored \cite{vorberg2023detection}. Previous research \cite{xia2021effective} targeting the pancreas has demonstrated that deep learning methodologies are capable of discerning nuanced textural and morphological alterations in NCT scans, which may elude even the observation of human experts. However, the feasibility of PE identification through NCT scans is still an open question, primarily due to the low contrast differentiation between the embolism and surrounding pulmonary vessels on NCT scans, compounded by the diverse morphological presentations of embolisms, which intensify this identification challenge.

To tackle the aforementioned issues and leverage dual-phase knowledge, we propose a novel Cross-Phase Mutual learNing framework (CPMN) for PE identification on NCT scans. In this work, the identification task is decoupled into classification and segmentation to improve the interpretability of identification with more supporting information. Our primary contributions can be articulated as follows: (1) The developed novel mutual learning framework CPMN unifies PE classification and segmentation tasks across dual-phase (CTPA and NCT scans), which can 
foster knowledge transferring from CTPA to NCT, thereby enhancing the performance of the model on NCT scans. (2) The presented Inter-Feature Alignment (IFA) strategy through an affinity graph captures pair-wise spatial feature similarities, guided by connection range and granularity parameters to enhance spatial contiguity and facilitate mutual learning transfer from the CTPA- to the NCT-pathway network. (3) The Intra-Feature Discrepancy (IFD) strategy realized through the designed dense center loss engenders a sharper demarcation within the feature space, facilitating precise segmentation of PE against complex backgrounds for each single-pathway network. (4) A large-scale dual-phase dataset containing 334 PE patients and 1,105 normal subjects has been established. The proposed CPMN achieves the leading identification performance, which is 95.4\%, 99.6\%, and 78.5\% in patient-level sensitivity, specificity, and segmentation dice on NCT scans, indicating the potential of our approach as a robust and precise tool for PE identification in clinical practice.


\section{Methodology}

\subsubsection{Problem Formulation.} In the training stage, given a set of pair-wise data, namely NCT and CTPA volume, the entire dataset is denoted by $S = \{(X^n_i, X^c_i,$ $ Y_i, M_i)| i = 1, 2,..., N\}$, where $X^n_i$ and $X^c_i$ are the $i$-th NCT and CTPA volume, with $Y_i$ being the voxel-wise segmentation label map of the same size as $X_i$ and $K$ channels. Here, $K = 2$ represents the background and embolism. $M_i \in \{0, 1\}$ is the classification label of the image, where 0 stands for “normal” and 1 for “PE”. In the testing stage, solely the NCT volume $X^n_i$ is provided, and the objective is to predict abnormality probability and generate an embolism mask.

\begin{figure}[t]
    \centering
    \includegraphics[width=\textwidth]{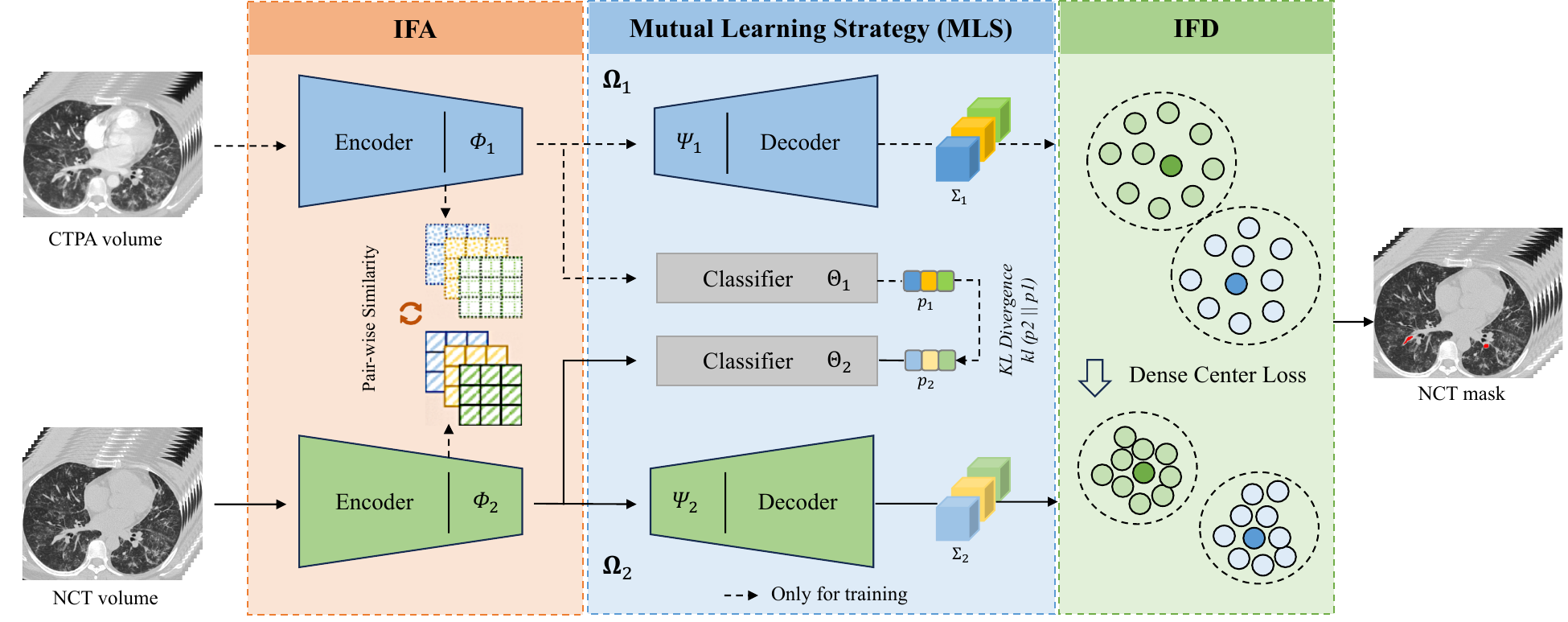}
\caption{Overview of our proposed Cross-Phase Mutual learNing framework (CPMN) that contains the CTPA-pathway network ($\mathbf{\Omega_{\text{1}}}$) and the NCT-pathway network ($\mathbf{\Omega_{\text{2}}}$). Each pathway network comprises an encoder-decoder pair ($\Phi_1/\Psi_1$, $\Phi_2/\Psi_2$) that extracts features from the corresponding volume. The presented Inter-Feature Alignment (IFA) strategy through an affinity graph captures pair-wise spatial feature similarities in the encoder. The predicted PE probabilities ($p_1$, $p_2$) are harmonized using KL divergence to align feature distributions without altering the CTPA-pathway network. The dense center loss is designed to refine the segmentation feature space ($\Sigma_1,  \Sigma_2$).}


    \label{fig:main_arch}
\end{figure}

\subsubsection{Mutual Learning Framework.}
In this section, we present our mutual learning framework designed for dual-phase medical image analysis, leveraging CTPA and NCT volumes. Our approach enhances the performance and generalization of the NCT-pathway network through a novel mutual learning strategy (MLS).

As shown in Fig \ref{fig:main_arch}, our proposed mutual learning framework is designed to simultaneously train two CNNs with distinct tasks,  classification, and segmentation. The CTPA-pathway network, denoted as $\mathbf{\Omega_{\text{1}}}$, is tasked with classification and segmenting PE in CTPA volumes. Conversely, the NCT-pathway network, $\mathbf{\Omega_{\text{2}}}$, operates on NCT volumes. We choose U-Net \cite{Ronneberger2015UNetCN}  3D version with EfficientNet-B0 \cite{Tan2019EfficientNetRM} 3D version as the encoder for segmentation task and add an auxiliary classifier $\Theta_{i, i \in \{1, 2\}}$ with architecture $\text{avg\_pool} \rightarrow \text{FC layer} \rightarrow \text{Relu} \rightarrow \text{FC layer}$, after encoder. The architecture is the same for both pathways.

Both networks are trained in parallel, as well as leveraging the MLS that fosters knowledge transfer from  $\mathbf{\Omega_{\text{1}}}$ to $\mathbf{\Omega_{\text{2}}}$. This is achieved by minimizing a divergence loss that aligns the predictive distributions of the two networks. We follow \cite{Zhang2017DeepML} employ the Kullback-Leibler (KL) divergence as a measure of the discrepancy between the output logits of $\boldsymbol{p}_{\text{1}}$ and $\boldsymbol{p}_{\text{2}}$, formulated as:
\begin{equation}
D_{K L}\left(\boldsymbol{p}_2 \| \boldsymbol{p}_1\right)= \sum_{i=1}^N \sum_{m=0}^1 p_2^m\left(\boldsymbol{x}_i\right) \log \frac{p_2^m\left(\boldsymbol{x}_i\right)}{p_1^m\left(\boldsymbol{x}_i\right)},
\mathcal{L_{\text{KL}}} =  D_{KL}\left( \boldsymbol{p}_{\text{2}}\,||\, \boldsymbol{p}_{\text{1}}\right)
\end{equation}
where $\boldsymbol{p}_{\text{1}}$ and $\boldsymbol{p}_{\text{2}}$ represent softmax probabilities from two classification heads ($\Theta
_1$, $\Theta_2$) given input $X_i^n$ and $X_i^c$ , respectively. By minimizing $\mathcal{L_{\text{KL}}}$, we encourage $\mathbf{\Omega_{\text{2}}}$ to adapt its predicted classification distribution towards that of $\mathbf{\Omega_{\text{1}}}$.

\subsubsection{Inter-Feature  Alignment (IFA).}
In our mutual learning framework, akin to the pair-wise Markov random field approach for enhancing spatial labeling contiguity, we focus on similarity among spatial features from the CTPA-pathway network $\mathbf{\Omega_{\text{2}}}$ to the NCT-pathway network $\mathbf{\Omega_{\text{1}}}$. Inspired by \cite{Liu2019StructuredKD}, an affinity graph is built to encapsulate this relationship. This graph is parameterized by connection range $\alpha$ and granularity $\beta$, optimizing the graph's resolution and the fidelity of spatial relations captured. The affinity graph, with $\frac{W' \times H'}{\beta}$ nodes and $\frac{W' \times H'}{\beta} \times \alpha$ connections, serve as a dynamic representation of spatial correlations, enhancing the mutual learning process between the networks $\mathbf{\Omega_{\text{1}}}$ and $\mathbf{\Omega_{\text{2}}}$.

To quantify the knowledge transfer between them and foster the mutual learning process, we introduce a pair-wise similarity distillation loss, integrating the squared differences of pair-wise similarities with a similarity term to measure the alignment between the networks' feature map:

\begin{equation}
\label{eq:l_align}
\mathcal{L}_{\text{alig}} = \frac{1}{Z} \sum_{i \in \mathcal{R'}} \sum_{j \in \alpha}  (a^{\mathbf{\Omega_{\text{1}}}}_{ij} - a^{\mathbf{\Omega_{\text{2}}}}_{ij})^2  , \mathcal{R'} \in \{1,2,3, ..., \frac{W' \times H'}{\beta} \}
\end{equation}
where $Z = W' \times H' \times \alpha$ serves as a normalization factor and $a^{\mathbf{\Omega_{\text{1}}}}_{ij}$ and $a^{\mathbf{\Omega_{\text{2}}}}_{ij}$ denote the similarity between the $i$-th and $j$-th nodes computed by networks $\mathbf{\Omega_{\text{1}}}$ and $\mathbf{\Omega_{\text{2}}}$, respectively. And similarity between two nodes is computed from the aggregated features $\mathbf{f}_i$ and $\mathbf{f}_j$ as  $a_{ij} = \frac{\mathbf{f}_i^\top \mathbf{f}_j}{\|\mathbf{f}_i\|_2 \|\mathbf{f}_j\|_2}$. In our
implementation, we use average pooling to aggregate $\beta \times C$ features in one node to be $1 \times C$. In the training process,  since we only want the features of NCT-pathway network  $\mathbf{\Omega_{\text{2}}}$ getting closer to the CTPA-pathway network $\mathbf{\Omega_{\text{1}}}$, only the parameters of the  $\mathbf{\Omega_{\text{2}}}$ is updated through $\mathcal{L}_{\text{alig}}$.

\subsubsection{Intra-Feature  Discrepancy (IFD).}

To enhance our segmentation model's capability to distinguish discriminative features between the background and the pulmonary embolism, we propose an IFD strategy that is based on designed dense center loss derived from center loss \cite{Wen2016centerloss}, traditionally used in classification tasks. In each training iteration, the centers are computed as the centroid features of the pixels belonging to the corresponding class in our segmentation mask. This modification to the center loss method, denoted as $\mathcal{L}_{\text{disc}}$, is defined as follows:

\begin{equation}
\mathcal{L}_{disc}=\frac{1}{2} \sum_{k=0}^{k=1}\left\|\boldsymbol{x}_k-\boldsymbol{c}_{k}\right\|_2^2,  
\frac{\partial \mathcal{L}_{disc}}{\partial \boldsymbol{x}_k}=\boldsymbol{x}_k-\boldsymbol{c}_{k} 
\end{equation}
where $\mathbb{I}$ is the indicator function,  $\boldsymbol{x}_k$  is the feature of the $k$-th pixel belonging to class $k$, and  $\boldsymbol{c}_{k}$ denotes the $k$-th class center of deep features. The update function for centers $\boldsymbol{c}_k$ is depicted as: 

\begin{equation}
\Delta \boldsymbol{c}_k=\frac{\sum_{j=0}^{j=1}\mathbb{I} \left(j=k\right) \cdot\left(\boldsymbol{c}_k-\boldsymbol{x}_j\right)}{1+\sum_{j=0}^{j=1} \mathbb{I}\left(k=j\right)}
\end{equation}
This approach enables our networks to effectively learn compact and separate clusters in the feature space for each class, a critical aspect of segmentation.

\subsubsection{Learning and Optimization.}
The total loss $\mathcal{L_{\text{Total}}}$ of CPMN is defined as:

\begin{equation}
    \mathcal{L_{\text{Total}}} = \mathcal{L_{\text{clas}}} + \mathcal{L_{\text{seg}}} + \lambda_1 \mathcal{L_{\text{KL}}} +\lambda_2 \mathcal{L_{\text{alig}}} + \lambda_3 \mathcal{L_{\text{disc}}}  
\end{equation}
where $\lambda_1, \lambda_2$, and $\lambda_3$ are set to 0.25, 10, and 0.1, making these loss value ranges comparable. The classification loss  $\mathcal{L_{\text{clas}}}$  is binary cross-entropy loss, and the segmentation loss  $\mathcal{L_{\text{seg}}}$ is optimized through focal loss \cite{Lin2017FocalLF} to address the class imbalance ratio between the pulmonary embolism area and the background.

\section{Experiment}

\subsection{Datasets}
\textbf{In-House Dataset:} We establish a large-scale dual-phase dataset (ALD-PE) containing 334 PE patients and 1,105 normal subjects from a cooperative hospital between the years 2019 and 2022. Each case encompasses a CTPA scan in conjunction with the corresponding NCT phase. We use the latest patients in 2022 as a hold-out test set, resulting in a training set of 269 PE patients and 881 normal subjects, and a test set of 65 PE patients and 224 normal subjects. We randomly selected 20\% of the training data as an internal validation set. LapIRN \cite{mok2021conditional} is employed to register the CTPA phase to the NCT phase, and then invite an experienced radiologist to annotate labels on the CTPA phase using CTLabeler \cite{wang2023cascaded}. The segmentation mask and the class label are annotated based on radiology reports and clinical records. 
\textbf{Public Benchmark:} The FUMPE dataset \cite{masoudi2018new} is one of the largest publicly available datasets in this field containing 8,792 CTPA images obtained from 35 patients. The partitioning of the training and test datasets aligns with the methodology delineated in prior research \cite{chen2024scunet++}.



\subsection{Implementation Details}

We developed our segmentation models using PyTorch, with experiments conducted on two NVIDIA A100 GPUs. We set the training batch size to 6, with transformations including random flips and rotations with  10\% probabilities, spatial padding, and random cropping to a uniform size of \(224 \times 224 \times 96\). During the inference stage, we use sliding-window inference with patch size \(224 \times 224 \times 96\), and the center patch is cropped with the same size as the input for the classification head. The Adam \cite{Kingma2014AdamAM} optimizer, with a learning rate of \(0.001\), is paired with a Cosine Annealing learning \cite{Loshchilov2016SGDRSG} rate scheduler that strategically modulates the learning rate over the training epochs, with the minimum rate set at \(1 \times 10^{-5}\). 

\subsection{Evaluation Metrics and Reader Study}
For the binary classification task, model performance is evaluated using the area under the Receiver Operating Characteristic curve (AUC), sensitivity (Sens.), and specificity (Spec.). For the segmentation task, the dice coefficient is utilized to assess model performance. 
A reader study was carried out involving three radiologists in cardiopulmonary imaging: an expert radiologist (12 years experience), a senior radiologist (8 years experience), and a junior radiologist (3 years experience). The readers were given 289 non-contrast CT scans from the test set and asked to provide a binary decision for each scan, determining the presence or absence of PE. They conducted their evaluations without access to any patient information or medical records. Additionally, readers were apprised that the dataset could exhibit a higher incidence of PE cases compared to the typical prevalence encountered in routine screenings, but the exact distribution of case types was not revealed to them. Utilizing the ITK-SNAP software \cite{yushkevich2006user}, the radiologists interpreted the CT scans, free from any time limitations.

\subsection{Results}
\subsubsection{Ablation Study.}

To verify the contribution of each component, the ablation study is carried out, and the results are reported in Table \ref{tab:ablation_study}. As for single-phase, our baseline model achieves 96.9\%, 99.6\%, 0.996, and 79.9\% in patient-level sensitivity, specificity, AUC, and segmentation dice on CTPA scans, while realizing 84.6\%, 97.8\%, 0.973, and 68.8\% on NCT scans. In the context of dual-phase analysis, our attention is exclusively dedicated to quantifying the performance enhancement conferred by each component on NCT scans. 
(1) \textbf{MLS:} The results demonstrate that the introduced MLS yields 7.7\% and 1.3\% improvement in sensitivity and specificity, which proves that MLS can synergy the strengths of both phases to enhance predictive performance on NCT scans. (2) \textbf{IFA:} Quantitative results show that the presented IFA strategy increases the segmentation dice from 70.1\% to 75.7\% while maintaining the classification performance, which indicates the effectiveness of the IFA strategy by constraint of pair-wise spatial feature similarities. (3) \textbf{IFD:} The results show that the designed IFD strategy can further improve the segmentation dice to 78.5\%. Concurrently, there is a notable enhancement in patient-level sensitivity by 3.1\%, culminating at 95.4\%. It proves the importance of the designed IFD strategy.

\subsubsection{Comparison with Literature.}

\begin{table}[t]
    \centering
    \setlength{\tabcolsep}{1.65mm}
    \caption{Ablation study on the test set of ALD-PE dataset. MLS: mutual learning strategy. IFA: inter-feature alignment. IFD: Intra-feature discrepancy. Sens.: Sensitivity. Spec.: Specificity. $^{\dagger}$: $p < 0.05$ for permutation test ((3) vs. NCT model and (2)). $^{\ddagger}$: $p < 0.05$ for DeLong test ((3) vs. NCT model). $^\bigtriangleup $: \textit{For the dual-phase, only the performance of the model on NCT scans is reported here.} }

    \begin{tabular}{cccccc}
    \toprule
    \multirow{2}{*}{Phase} & \multirow{2}{*}{Method} & \multicolumn{3}{c}{Classification} & \multicolumn{1}{c}{Segmentation}\\
    \cmidrule(lr){3-5} \cmidrule(lr){6-6}
     &    &   Sens. (\%)     & Spec. (\%)           & AUC          & Dice (\%)   \\ 
    \midrule
    \multirow{2}{*}{$Single$} &  CTPA model & 96.9 & 99.6 & 0.996 & 79.9   \\
     & NCT model &  84.6 & 97.8 & 0.973 &   68.8   \\
    \midrule
    \multirow{3}{*}{$Dual^\bigtriangleup$} & + MLS$^{(1)}$ & 92.3  &  99.1   & 0.989  &  70.1   \\
     & + MLS + IFA$^{(2)}$ & 92.3  & 99.1  & 0.988  &  75.7   \\
     & + MLS + IFA + IFD$^{(3)}$ & \textbf{95.4}$^{\dagger}$  & \textbf{99.6}$^{\dagger}$ &  \textbf{0.990}$^{\ddagger}$ &  \textbf{78.5} \\
    \bottomrule
    \end{tabular}
\label{tab:ablation_study}
\vspace{-1.5em}
\end{table}

\begin{table}[t]
    \centering
    \setlength{\tabcolsep}{3.3mm}
    \caption{Comparison with radiologists and literature on the test set of ALD-PE dataset for NCT scans. Sens.: Sensitivity. Spec.: Specificity. $^\dagger$: $p < 0.05$ for permutation test (CPMN vs. nnFormer-Joint and radiologist experts). $^*$: $p < 0.05$ for the DeLong test (CPMN vs. nnFormer-Joint). } 
    \begin{tabular}{ccccc}
    \toprule
    \multirow{2}{*}{Method} & \multicolumn{3}{c}{Classification} & \multicolumn{1}{c}{Segmentation}\\
    \cmidrule(lr){2-4} \cmidrule(lr){5-5}
      &   Sens. (\%)     & Spec. (\%)           & AUC          & Dice (\%)   \\ 
    \midrule
    Mean of radiologists & 38.5 & 78.6 & - & -   \\
    \midrule
    DML \cite{Zhang2017DeepML} & 89.2 & 97.8 & 0.986 & - \\
    nnU-Net-Joint \cite{isensee2021nnu} & 87.7 & 98.2 & 0.969 & 72.4  \\
    Mask2Former-Joint \cite{cheng2022masked} & 86.2 & 98.7 & 0.955 &  70.6    \\
    nnFormer-Joint \cite{zhou2023nnformer} & 89.2 & 98.7 & 0.976 &  73.2  \\
    \midrule
    \textbf{CPMN} & \textbf{95.4}$^\dagger$  & \textbf{99.6}$^\dagger$ & \textbf{0.990}$^*$  & \textbf{78.5}  \\
    \bottomrule
    \end{tabular}
\label{tab:compare}
\vspace{-1em}
\end{table}

\begin{figure}[t]
    \centering
    \includegraphics[width=\textwidth]{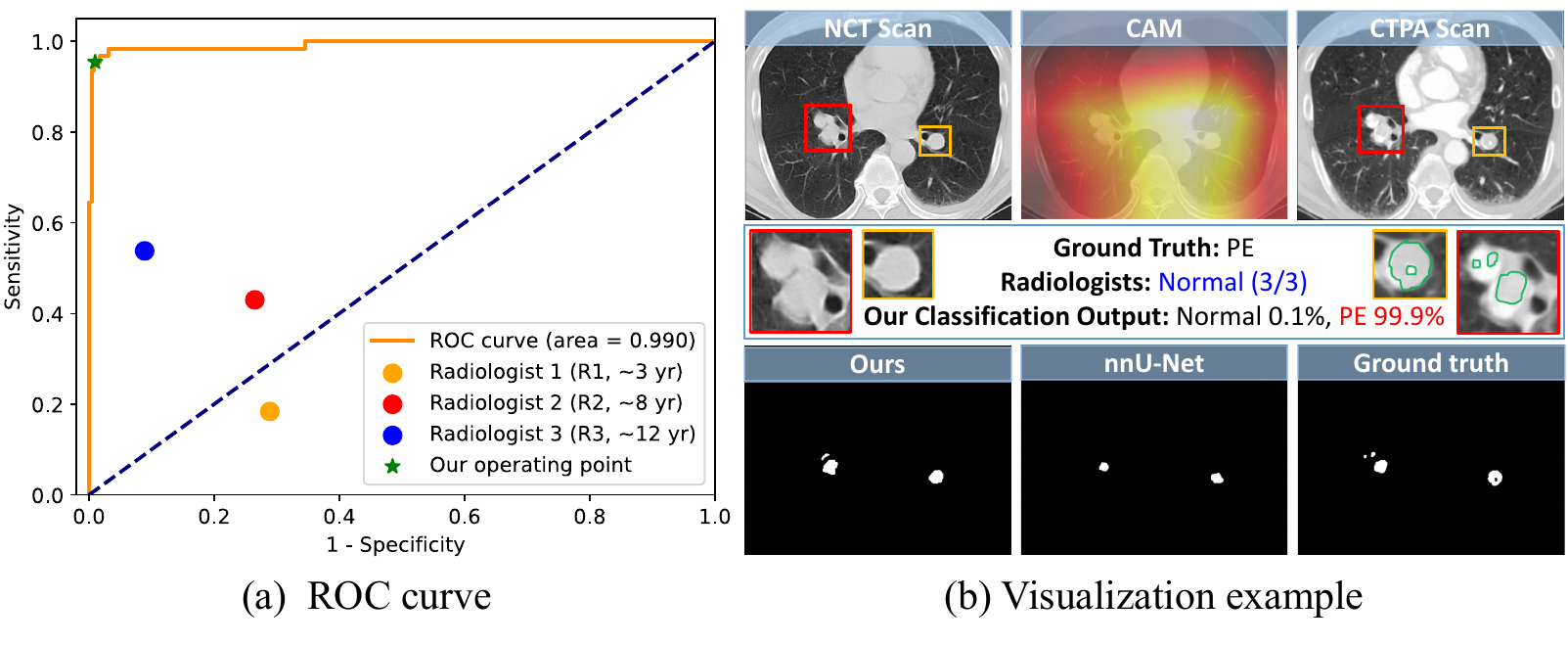}
    \caption{(a) ROC curve for our model versus three radiologists on the hold-out test set (n = 289) for binary classification. (b) Visualization example in the test set. This PE case is miss-detected by three radiologists but our model succeeds in locating the embolism by CAM \cite{zhou2016learning} and predicted mask. Green contours represent the regions of embolism (best viewed in color).}
    \label{fig:vis}
\vspace{-1em}
\end{figure}

To evaluate the effectiveness of our proposed CPMN, we conduct a comparison with various state-of-the-art methods on two different datasets. (1) \textbf{ALD-PE:} Table \ref{tab:compare} presents a comparative analysis of our proposed CPMN with four baselines. The first baseline is DML \cite{Zhang2017DeepML} based on mutual learning. The other three baselines (denoted as “-Joint”) integrate a CNN classification head into each network and are trained in an end-to-end manner. Quantitative results show that our proposed CPMN achieves the leading classification and segmentation performance, particularly in sensitivity and segmentation dice. Qualitative results, as shown in Fig. \ref{fig:vis}(b), demonstrate that CPMN achieves more robust segmentation results.
(2) \textbf{FUMPE:} We assess the efficacy of our training framework for segmentation tasks on single-phase data streams. To conduct this evaluation, we modify the CTPA-pathway network by removing the auxiliary classification head and using the 2D-EfficientNet-B0 and 2D-U-Net architecture. All other components of the CTPA-pathway network remain unchanged. This approach achieves a dice coefficient of 77.4\%, surpassing the performance of a recent dedicated model (ResD-UNet \cite{Yuan2021ResDUnetRA}) tailored for PE identification, which achieves 76.5\%. The results demonstrate that our introduced single-pathway network is robust and effective. More importantly, the comparison results further substantiate that the advancements in identification on NCT scans are not solely due to the choice of a powerful backbone but rather the effectiveness of the proposed CPMN itself.

\vspace{-1em}
\subsubsection{Comparison with Radiologists.}

As shown in Fig. \ref{fig:vis}(a), our proposed CPMN’s ROC curve is superior to the performance of three radiologists. The model achieves a patient-level sensitivity of 95.4\% in PE identification, which significantly exceeds that of radiologists (18.5\%, 43.1\%, and 53.8\%) while maintaining a high specificity of 99.6\%. A visual example is presented in Fig. \ref{fig:vis}(b), which is miss-detected by three radiologists, whereas classified and localized precisely by  CPMN. More importantly, the extra information about predicted embolism masks and CAM \cite{zhou2016learning} generated from the classification head improve the interpretability of identification.

\section{Conclusion}
In this work, a novel Cross-Phase Mutual learNing framework (CPMN) has been proposed to facilitate knowledge transfer from CTPA to NCT, thereby enhancing performance for PE identification on NCT scans. Additionally, the framework provides outputs of CAM and embolism masks for improved clinical interpretability. The comprehensive evaluation demonstrates that our approach outperforms strong baselines and
experienced radiologists,  highlighting the potential of our approach as a robust and precise tool for PE identification in real clinical environments.

\subsubsection{Acknowledgement.}
This work was supported by the Technical Innovation Key Project of Zhejiang Province (2024C03023) to H.Z.

\subsubsection{Disclosure of Interests.}
The authors declare no competing interests.
%
%
%
\bibliographystyle{splncs04}
\bibliography{Paper-2986}
%




\end{document}